\documentclass[]{eas}

\usepackage{amssymb,graphicx}
\usepackage{natbib}

\newcommand{\avg}[1]{\langle#1\,\rangle}  
\newcommand{\cmcube}{\ensuremath{\,\mathrm{cm^{-3}}}} 
\newcommand{\degree}{\ensuremath{^{\circ}}}

\newcommand{\HI}{\ensuremath{\mathrm{HI}}}

\newcommand{\n}{\ensuremath{\,n_\mathrm{e}}} 
 
\newcommand{\nH}{\ensuremath{\,n_\mathrm{H\,\scriptstyle I}}}

\newcommand{\pc}{\ensuremath{\,\mathrm{pc}}}
\begin{document}
%
\title{Density PDFs of diffuse gas in the Milky Way}
\runningtitle{Density PDFS of diffuse gas}
\author{Elly M. Berkhuijsen}
\address{Max-Planck-Institut f\"ur Radioastronomie, Auf dem H\"ugel 69, 53121 Bonn, Germany.}
\author{Andrew Fletcher}
\address{School of Mathematics and Statistics, Newcastle University, NE1 7RU, U.K.}
%
\begin{abstract}
The probability distribution functions (PDFs) of the average densities of the diffuse ionized gas (DIG) and the diffuse atomic gas are close to lognormal, especially when lines of sight at $|b|<5\degree$ and $|b|\ge5\degree$ are considered separately. Our results provide strong support for the existence of a lognormal density PDF in the diffuse ISM, consistent with a turbulent origin of density structure in the diffuse gas.
\end{abstract}
\maketitle
\section{Introduction}
Simulations of the interstellar medium (ISM) have shown that, if isothermal
turbulence is shaping the structure of the medium, the density distribution
becomes lognormal \citep{Elmegreen:2004, Avillez:2005, Wada:2007}. The latter
authors found that for a large enough volume, and for a long enough simulation
run, the physical processes causing the density variations in the ISM in a
galactic disc can be regarded as random and independent events. Therefore, the
PDF of log(density) becomes Gaussian and the density PDF lognormal. The medium
is inhomogeneous on a local scale, but in a quasi-steady state on a global
scale.

Little observational evidence exists to test the results of the 
simulations. To date lognormal PDFs have been derived for the $\HI$ column densities in the LMC  \citep{Wada:2000}, for emission measures perpendicular to the Galactic plane of the DIG in 
the Milky Way \citep{Hill:2007, Hill:2008}, and for the local volume density of dust near stars within 400\pc\ of the Sun \citep{Gaustad:1993}. Here we 
present the PDFs of average volume densities of the DIG and the 
diffuse atomic gas in the Milky Way, and show that they are consistent 
with lognormal distributions as well. 

We investigated the density PDFs of the DIG 
and of diffuse atomic hydrogen gas ($\HI$) in the solar neighbourhood, using the dispersion measures of 34 pulsars at known distances \citep{Berkhuijsen:2008} and $\HI$ absorption lines towards 375 stars \citep{Diplas:1994a, Diplas:1994b}. Further details and an extensive discussion of some of the results presented here are given in \citet{Berkhuijsen:2008a}.

\section{Results and discussion}
\label{sec:results}

\begin{figure}
\center
\includegraphics[width=0.61\textwidth]{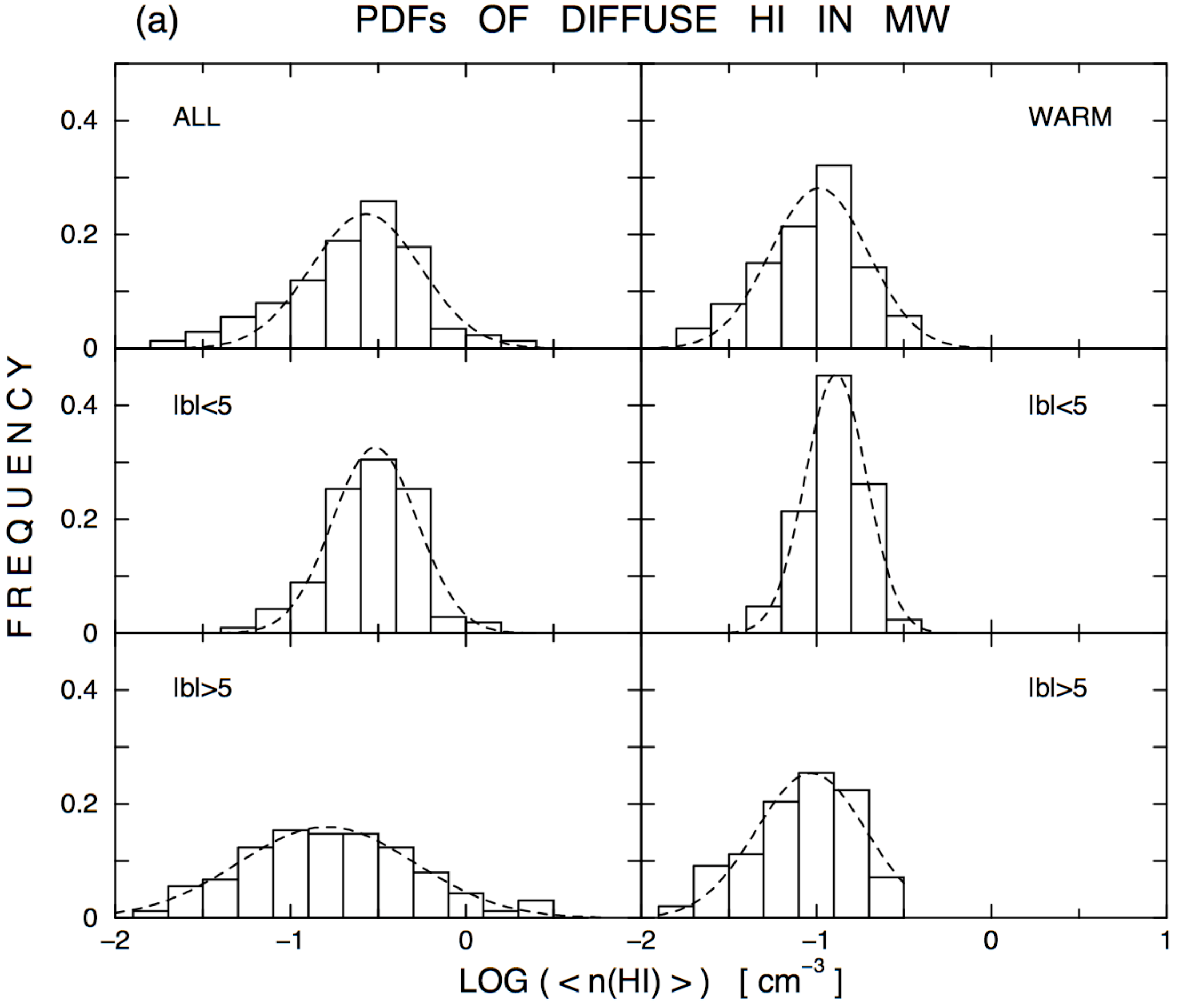}
\hfill
\includegraphics[width=0.34\textwidth]{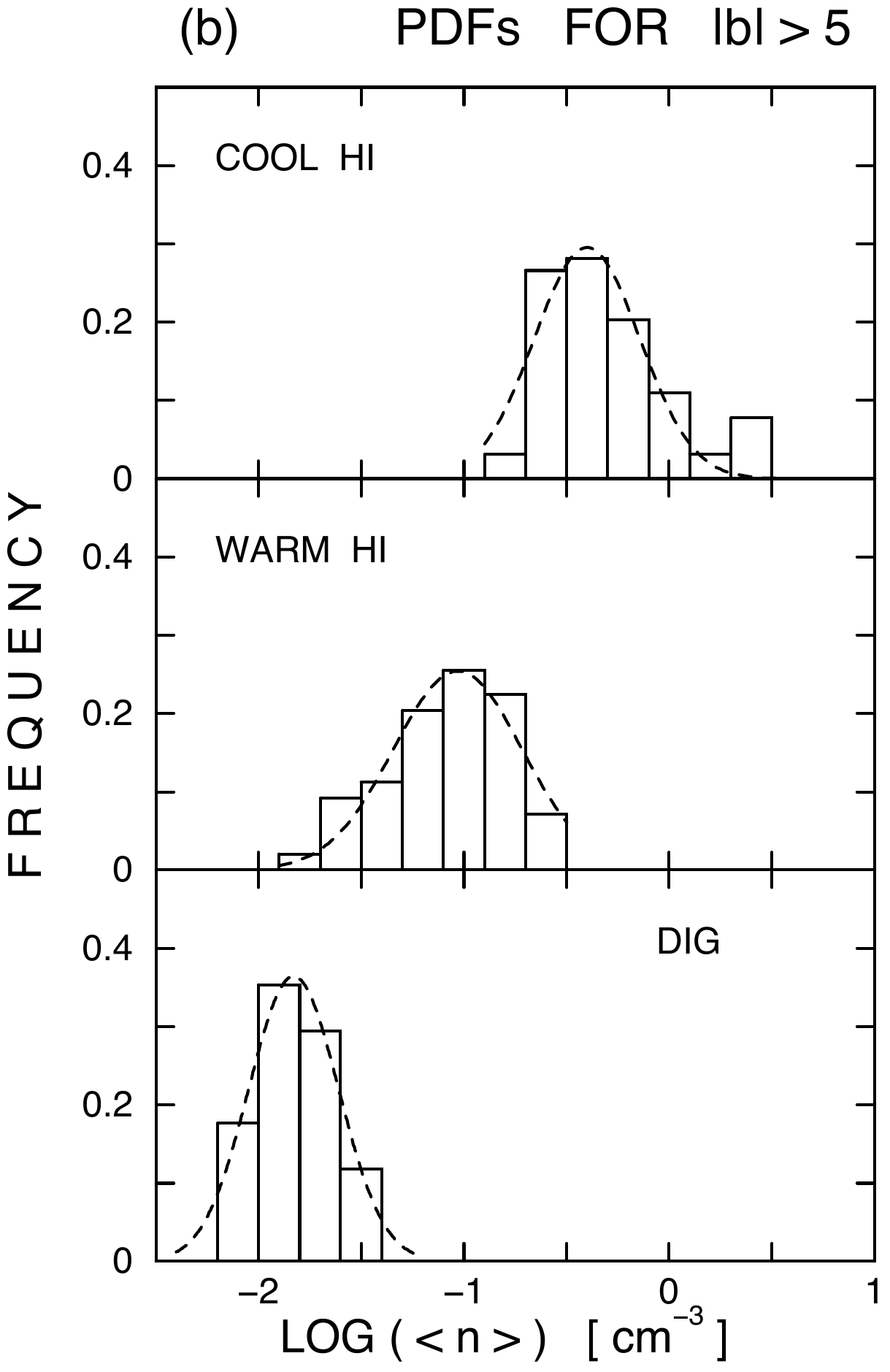}
\caption{\textbf{(a)} Six panels showing PDFs of $\HI$. Left panels: PDFs of all diffuse $\HI$, right panels: PDFs of only warm $\HI$. \textbf{(b)} Three panels showing PDFs at high latitude ($|b|\ge 5\degree$) for two $\HI$ phases and the diffuse ionized gas.}
\label{fig}
\end{figure}

\begin{table*}
\caption{Lognormal fits to the PDFs of $\avg{\nH}$ and $\avg{\n}$. The fitted function is \hspace{1cm}$Y=(\sqrt{2\pi}\sigma)^{-1}\exp[-(\log_{10} \avg{n} - \mu)^2/2\sigma^2]$, where $Y$ is the normalised number of pulsars or stars per bin, and the reduced chi-square value for each fit is given.}
\begin{center}
\begin{tabular}{lrrrrr}
\hline
 & & \multicolumn{2}{c}{Position of maximum} & Dispersion & \\
Area & $N$ & $\mu$ & $\avg{n}$ & $\sigma$ & $\chi^2$\\
\hline
All stars & $375$ & $-0.57\pm 0.03$ & $0.27\pm 0.02$ & $0.34\pm 0.03$ & $6.0$\\
  $|b|<5\degree$ & $213$ & $-0.52\pm0.02$ & $0.30\pm 0.02$ & $0.24\pm 0.02$ & $2.7$\\
  $|b|\ge 5\degree$ & $162$ & $-0.80\pm 0.02$ & $0.16\pm 0.01$ & $0.50\pm 0.02$ & $0.4$\\
  & & & & & \\
Cool $\HI$ $|b|\ge 5\degree$ & $64$ & $-0.40\pm0.05$ & $0.40\pm0.04$ & $0.27\pm0.03$ & $1.3$ \\
  & & & & & \\
Warm $\HI$ & $140$ & $-0.98\pm 0.03$ & $0.10\pm 0.01$ & $0.28\pm 0.03$ & $2.5$\\
  $|b|<5\degree$ & $42$ & $-0.89\pm 0.01$ & $0.13\pm 0.01$ & $0.18\pm 0.01$ & $0.1$\\
  $|b|\ge 5\degree$ & $98$ & $-1.07\pm 0.03$ & $0.09\pm 0.01$ & $0.33\pm 0.03$ & $1.0$\\
  & & & & & \\
DIG $|b|\ge 5\degree$ & $34$ & $-1.83\pm0.02$ & $0.015\pm0.001$ & $0.22\pm0.02$ & $0.3$ \\
\hline\noalign\\  
\end{tabular}
\end{center}
\label{table}
\end{table*}

In Fig.~\ref{fig}(a)(left panels) we present the PDFs of the average volume density of
$\HI$, $\avg{\nH}$, for the full sample of 375 stars (top) and for the stars at low ($|b|<5\degree$) and high ($|b|\ge 5\degree$) latitudes (middle and bottom panels)
\citep{Diplas:1994a}. The PDF for all stars is approximately lognormal above $\mathrm{log}\avg{\nH} = -1$, but there is a clear excess at
lower densities reflected in the large reduced-$\chi^2$ statistic (see Table~\ref{table}). 
The PDFs for low and high $|b|$ have a lognormal shape but their maxima are shifted with respect to each other. The latter sample clearly causes the low-density excess in
Fig.~\ref{fig}(a)(top left). This is also the case for the PDFs for the lines of sight (LOS) towards stars probing warm $\HI$ \citep{Diplas:1994b} shown in Fig.~1(a) (right panels). The shifts between the maxima show the difference in mean density between low and high latitudes (see Table~\ref{table}). Note that all densities of the warm sample have $\avg{n}<0.3\cmcube$.  

The dispersions of the PDFs of the high $|b|$ samples are twice those of the low $|b|$
samples. A possible explanation is that the low $|b|$ LOS cross more turbulent cells than at high $|b|$ \citep{Vazquez-Semadeni:2001}.

Table~\ref{table} shows that the
dispersions of the PDFs of the warm $\HI$ are slightly smaller than for
the full sample. The combination of warm and cool (denser) gas in the full sample
increases the dispersion because the density range becomes larger. This is clearly visible in Fig.~\ref{fig}(b): addition of the samples at high $|b|$ of cool $\HI$ (top) and warm $\HI$ (middle) gives the distribution for all high $|b|$ LOS shown in the bottom-left panel of Fig.~\ref{fig}(a). However, the mixture of cool and warm $\HI$ only partly explains the difference in dispersion of the full samples at low and high $|b|$, because this difference also exists for the samples of warm $\HI$.

In the bottom panel of Fig.~\ref{fig}(b) we show the PDF of DIG, which is also lognormal, below that of the warm $\HI$ at the same $|b|$. The temperatures of the two components are
similar and if the ionized and atomic gas are well mixed, one would expect
their dispersions to be the same. However, the dispersion of the DIG sample is about $30$ per cent smaller than that of the warm $\HI$ sample (see Table~\ref{table}). A plausible explanation for the difference, which is also consistent with the higher density of the maximum in the diffuse $\avg{\nH}$ PDF, is that low density regions are more readily ionized than higher density gas and that the average degree of ionization of the diffuse gas is substantially lower than $50$ per cent. We estimate the degree of ionization to be about $14$ per cent, using the densities for $\avg{\n}$ and $\avg{\nH}$ in Table~\ref{table}. Alternatively, the DIG could have a higher mean temperature than the warm $\HI$ but with a smaller temperature range; in the simulations of \citet{Avillez:2005} high temperature gas indeed has a lower median density and smaller dispersion.

\section{Conclusions}

The results presented in Section~\ref{sec:results} lead us to the following conclusions:

\begin{enumerate}
\item The density PDF of the diffuse ISM is approximately lognormal, but there is a significant low density excess in the best-fit distribution.
\item The density PDFs of the diffuse ISM in the disk ($|b|<5\degree$) and away from the disk ($|b|\ge 5\degree$) are lognormal, but the positions of their maxima and the dispersions differ. These differences produce the low density excess in the PDF for all lines of sight.
\item Several effects seem to influence the shape of the PDF. An increase of the number of clouds/cells along the LOS causes a \emph{decrease} in the dispersion, whereas a mixture of cool (dense) and warm (less dense) gas causes an \emph{increase} in the dispersion.
\end{enumerate}

These results provide strong support for the existence 
of a lognormal density PDF in the diffuse (i.e. average densities of 
$n<1\cmcube$) ionized and neutral components of the ISM. In turn, the form of 
the PDFs is consistent with the small-scale structure of the diffuse ISM being controlled 
by turbulence.


\end{document}